%% file: main.tex
\documentclass{article}

\usepackage{arxiv}

\usepackage[utf8]{inputenc} % allow utf-8 input
\usepackage[T1]{fontenc}    % use 8-bit T1 fonts
\usepackage{hyperref}       % hyperlinks
\usepackage{url}            % simple URL typesetting
\usepackage{booktabs}       % professional-quality tables
\usepackage{amsfonts}       % blackboard math symbols
\usepackage{nicefrac}       % compact symbols for 1/2, etc.
\usepackage{microtype}      % microtypography
\usepackage{lipsum}
\usepackage{graphicx}
\graphicspath{{./images/}}

\input{preamble.tex}

\usepackage{colortbl}
\usepackage{xcolor}
\usepackage{float}
\usepackage{enumitem}
\usepackage{setspace}
\usepackage{tikz}
\usetikzlibrary{spy,backgrounds}
\usepackage{pifont}
\usepackage{xspace}
\usepackage{multirow}

\newcommand{\cmark}{\ding{51}}
\newcommand{\xmark}{\ding{55}}
\newcommand{\greencheck}{{\color{green}\cmark}\xspace}

\newcommand{\redcross}{{\color{red}\xmark}\xspace}

\newcommand{\orangecross}{{\color{orange}\xmark}\xspace}
\newcommand{\green}{\cellcolor{green!10}\greencheck}
\newcommand{\orange}{\cellcolor{yellow!10}\orangecross}
\newcommand{\red}{\cellcolor{red!10}\redcross}
\newcommand{\etal}{\textit{et al.}}

\title{End-to-End Hybrid Refractive-Diffractive Lens Design with Differentiable Ray-Wave Model}
\author{
Xinge Yang$^{1}$, Matheus Souza$^{1}$, Kunyi Wang$^{1,2}$, Praneeth Chakravarthula$^{3}$, Qiang Fu$^{1}$, Wolfgang Heidrich$^{1,*}$ \\ \\
KAUST$^{1}$, UBC$^{2}$, UNC$^{3}$, Corresponding Author$^{*}$\\
}

\begin{document}

\maketitle

\begin{abstract}
Hybrid refractive-diffractive lenses combine the light efficiency of refractive lenses with the information encoding power of diffractive optical elements (DOE), showing great potential as the next generation of imaging systems. However, accurately simulating such hybrid designs is generally difficult, and in particular, there are no existing differentiable image formation models for hybrid lenses with sufficient accuracy. 

In this work, we propose a new hybrid ray-tracing and wave-propagation (ray-wave) model for accurate simulation of both optical aberrations and diffractive phase modulation, where the DOE is placed between the last refractive surface and the image sensor, i.e. away from the Fourier plane that is often used as a DOE position. The proposed ray-wave model is fully differentiable, enabling gradient back-propagation for end-to-end co-design of refractive-diffractive lens optimization and the image reconstruction network. We validate the accuracy of the proposed model by comparing the simulated point spread functions (PSFs) with theoretical results, as well as simulation experiments that show our model to be more accurate than solutions implemented in commercial software packages like Zemax. We demonstrate the effectiveness of the proposed model through real-world experiments and show significant improvements in both aberration correction and extended depth-of-field (EDoF) imaging. We believe the proposed model will motivate further investigation into a wide range of applications in computational imaging, computational photography, and advanced optical design. Code will be released upon publication.
\end{abstract}

\input{1_intro.tex}
\input{2_related_works.tex}
\input{3_methods.tex}
\input{4_verification.tex}
\input{5_simu_results.tex}
\input{6_real_results.tex}

\input{7_conclusion.tex}

\bibliographystyle{unsrt}  
\bibliography{ref}  %%% Remove comment to use the external .bib file (using bibtex).

\end{document}

%% file: preamble.tex
\usepackage{graphicx}
\usepackage{amsmath}
\usepackage{amssymb}
\usepackage{booktabs}
\usepackage{float}
\usepackage{stfloats}
\usepackage{tikz}
\usetikzlibrary{spy,backgrounds}
\usepackage{diagbox}
\usepackage{multirow}
\usepackage{wrapfig}
\usepackage{lipsum}
\usepackage{color, colortbl}
\usepackage{caption}
\usepackage{pifont}
\usepackage{enumitem}
\usepackage{adjustbox}

%% file: 1_intro.tex
\section{Introduction}

\begin{figure}[ht]
  \includegraphics[width=\textwidth]{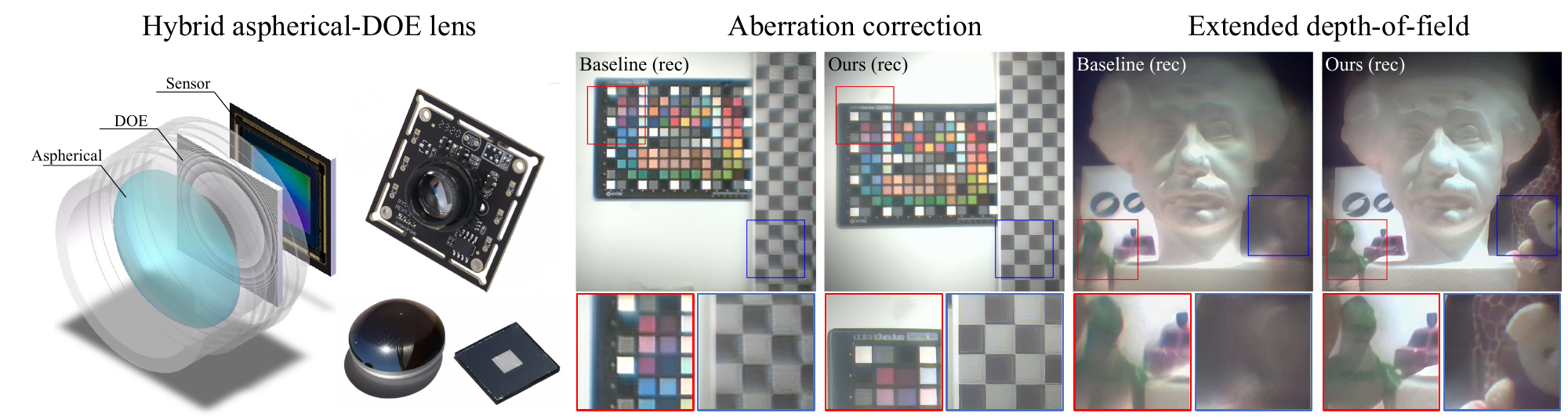}
  \caption{We propose a differentiable ray-wave imaging model that accurately simulates both aberrations and phase modulation while enabling end-to-end optimization for the hybrid refractive-diffractive lens and the image reconstruction network. To experimentally demonstrate the effectiveness of the proposed model, we present a hybrid aspherical-DOE lens prototype (left) and investigate two applications: aberration correction (middle) and large field-of-view extended depth-of-field imaging (right).}
  \label{fig:teaser}
\end{figure}

End-to-end optical design~\cite{sitzmann2018end,sun2020learning,wu2019phasecam3d,shi2022seeing,metzler2020deep,wetzstein2020inference} has demonstrated remarkable potential in scientific imaging~\cite{jeon2019compact,dun2020learned,baek2021single,li2022quantization,sun2020learning,metzler2020deep,shi2022seeing,tseng2021neural}, computer vision~\cite{wu2019phasecam3d,ikoma2021depth,cote2023differentiable,tseng2021differentiable,yang2023image}, and advanced optical system design~\cite{yang2023curriculum,zheng2023close}, surpassing classical lens design approaches. An end-to-end lens system comprises an optical encoder, such as a refractive lens, the diffractive optical element (DOE), and/or metasurface~\cite{tseng2021neural,chen2016review}, which captures information from the real world, and a neural network decoder that reconstructs the final output, which can be an image or other visual representation. The optics and the network are jointly optimized using gradient backpropagation to find the optimal computational optical system for a given task. Among existing end-to-end computational lenses, hybrid refractive-diffractive lenses~\cite{wang2016chromatic,flores2004achromatic} combine the advantages of both refractive and diffractive optics, showcasing the great potential for next-generation optical imaging systems.

However, no existing differentiable image formation model provides sufficient accuracy for hybrid refractive-diffractive lenses, posing a significant challenge for the end-to-end design of such systems. The commonly used paraxial wave optics model~\cite{goodman2005introduction} idealizes the refractive lens as a thin phase plate and neglects optical aberrations. The ray tracing model, employed in commercial optical design software such as Zemax~\cite{ZemaxManual} and in recent works~\cite{zhang2024centimeter,shih2024hybrid}, simplifies the diffractive element as local gratings and introduces an approximation for diffraction simulation~\cite{fischer2000optical}. However, both of these simplified optical models fail to accurately simulate aberrations and diffraction simultaneously, and they rely on strong assumptions to function.

In this paper, we propose a differentiable ray-tracing and wave-propagation (ray-wave) model for accurate simulation of hybrid lens systems. The ray-wave model can simulate both refractive optical aberrations and diffractive phase modulation in a differentiable manner, enabling end-to-end co-design of refractive lenses, DOEs, and neural networks. Specifically, we study a hybrid lens configuration where the DOE is placed between the refractive lenses and the image sensor. The proposed ray-wave model first performs coherent ray tracing~\cite{mout2018ray,chen2021optical} to accurately simulate the aberrated amplitude and phase map, followed by scalar Rayleigh-Sommerfeld diffraction~\cite{goodman2005introduction} to incorporate diffractive phase modulation. Free-space propagation to the image plane is then performed for point spread function (PSF) calculation. In contrast to existing imaging models, our ray-wave model combines the advantages of both ray tracing and wave optics, providing accurate simulation for optical aberrations and diffractive phase modulation.

First, we validate the proposed ray-wave model by comparing the PSF simulation results with both theoretical predictions and the ray tracing model. The ray-wave model provides accurate simulation results and is capable of handling discontinuous diffractive phase maps, whereas the ray tracing model fails to accurately capture real diffractive phenomena. Next, we design a compound hybrid refractive-diffractive lens integrated with an image reconstruction network for computational imaging. We then compare its performance with both a paraxial achromat design and a design generated using Zemax optical software. The proposed end-to-end hybrid lens design demonstrates superior image reconstruction quality compared to existing methods.

Furthermore, we demonstrate a hybrid aspherical-DOE lens prototype (Fig.~\ref{fig:teaser}) featuring a large field-of-view (FoV), compact form factor, and high image quality. To validate the effectiveness of our model, we investigate two real-world applications. First, we perform an end-to-end design of a DOE for computational aberration correction. Experimental results show that our hybrid lens successfully mitigates optical aberrations of the refractive component, producing high-quality images. Second, we design a DOE for large FoV extended depth-of-field (EDoF) imaging. Unlike existing DOE designs that idealize the refractive lens as a thin phase plate, our design explicitly accounts for lens aberrations. As a result, our DOE design significantly improves the reconstructed image quality, especially in off-axis regions where aberrations are more pronounced.

The main contributions of this paper can be summarized as:
\begin{itemize}[leftmargin=*]
\item We present a differentiable ray-wave model for hybrid refractive-diffractive lenses. The proposed model can accurately simulate both optical aberrations and diffractive phase modulation. It facilitates end-to-end optimization of refractive lenses, DOEs, and image reconstruction networks.
\item We demonstrate a hybrid aspherical-DOE lens prototype. Two real-world applications are investigated to validate the effectiveness of our model: computational aberration correction and large FoV EDoF imaging. Our hybrid lens achieves high-quality imaging performance with a large FoV and compact form factor.
\end{itemize}

%% file: 2_related_works.tex
\section{Related Works}

\subsection{End-to-End Optical Design}

Classical optical design~\cite{kingslake2012lens,kidger2001fundamental,fischer2000optical} optimizes the lens system for the best optical performance independently of the image processing algorithm. However, with the advancement of deep learning and computer vision, the intermediate raw captures of cameras can be post-processed by neural networks, eliminating the need for them to be the design objective. Following this idea, end-to-end optical design~\cite{sitzmann2018end,wu2019phasecam3d,wetzstein2020inference,sun2020learning,tseng2021neural}, jointly optimizes the optical system (typically containing both refractive and diffractive elements) and the image processing network for the best final output. In an end-to-end optical design pipeline, a differentiable image formation model is employed to simulate the raw captures, which are then fed into the image processing network. The optical system and the network can be jointly optimized using gradient backpropagation and deep learning algorithms. Recent research on end-to-end optical design has demonstrated outstanding performance in various applications, e.g., hyperspectral imaging~\cite{jeon2019compact,dun2020learned,baek2021single,li2022quantization}, high-dynamic-range imaging~\cite{sun2020learning,metzler2020deep}, seeing through obstructions~\cite{shi2022seeing}, depth estimation~\cite{wu2019phasecam3d,ikoma2021depth,chang2019deep}, object detection~\cite{cote2023differentiable,tseng2021differentiable}, and compact optical systems~\cite{tseng2021neural,chakravarthula2023thin}.

An accurate and differentiable image formation model is crucial for end-to-end optical design. Existing end-to-end design works typically use paraxial optical models~\cite{goodman2005introduction,sun2020learning,sitzmann2018end,shi2022seeing}, which idealize the refractive lens as a thin phase plate and neglect optical aberrations. However, this paraxial approximation is inaccurate and cannot optimize the refractive lens, limiting the designed optical systems to small fields of view and low aberration performance. Differentiable ray tracing~\cite{sun2021end,wang2022differentiable,yang2023curriculum,cote2023differentiable} is another widely used approach for the simulation of refractive lenses and can be extended to simulate diffractive surfaces by approximating them as local gratings~\cite{fischer2000optical}. Ray tracing for diffractive surfaces has been employed in both commercial optical design software, such as Zemax~\cite{ZemaxManual}, and recent research works~\cite{zhang2024centimeter,zhu2023metalens,shih2024hybrid}. However, while ray tracing can approximate the light propagation direction after diffractive surfaces, it cannot accurately represent real diffraction phenomena. Additionally, it requires the phase map of the diffractive surface to change slowly, whereas in end-to-end optical design, the DOE often has rapidly changing phase patterns. In this work, we propose a hybrid ray tracing and wave propagation (ray-wave) model for the accurate simulation of optical aberrations and phase modulation in hybrid refractive-diffractive lenses. This model combines ray tracing for aberration simulation with wave propagation for diffraction simulation.

\subsection{Hybrid Refractive-Diffractive Lens}

Hybrid refractive-diffractive lenses~\cite{stone1988hybrid,wang2016chromatic,flores2004achromatic} have showcased the great potential for next-generation optical imaging systems with a compact form factor, powerful information encoding capability, and outstanding light efficiency. Existing works on hybrid lenses can be summarized into two groups. In classical optical design, the DOE is primarily utilized for chromatic aberration correction with reduced physical size, given the reversed direction of dispersion in diffractive and refractive optics~\cite{wang2016chromatic,chen2018broadband,flores2004achromatic,stone1988hybrid}. For instance, Canon has revealed several patents for compact hybrid refractive-diffractive lenses~\cite{JP2011221510A}. Although the simulation and design process of these products are not publicly known, they use slowly varying phase patterns for which ray-tracing approaches work well. In more recent end-to-end optical design works~\cite{shi2022seeing,sun2020learning,wu2019phasecam3d}, where the DOE is typically designed with more complex phase patterns, sometimes even discontinuous with pixel-wise optimization, the ray-tracing models fail. Although existing end-to-end optical design works have demonstrated remarkable performance in numerous applications, the idealization of the DOE as a thin phase plate under the paraxial approximation is inaccurate. This limitation restricts the designed DOE to function only for low aberrations and small FoV when combined with refractive lenses. Additionally, the paraxial optics model can not optimize the refractive lens. So far, there is no existing work that can accurately simulate and optimize refractive lenses with complex and discontinuous diffractive surfaces.

%% file: 3_methods.tex
\begin{figure*}[thp]
    \centering
    \includegraphics[width=\textwidth]{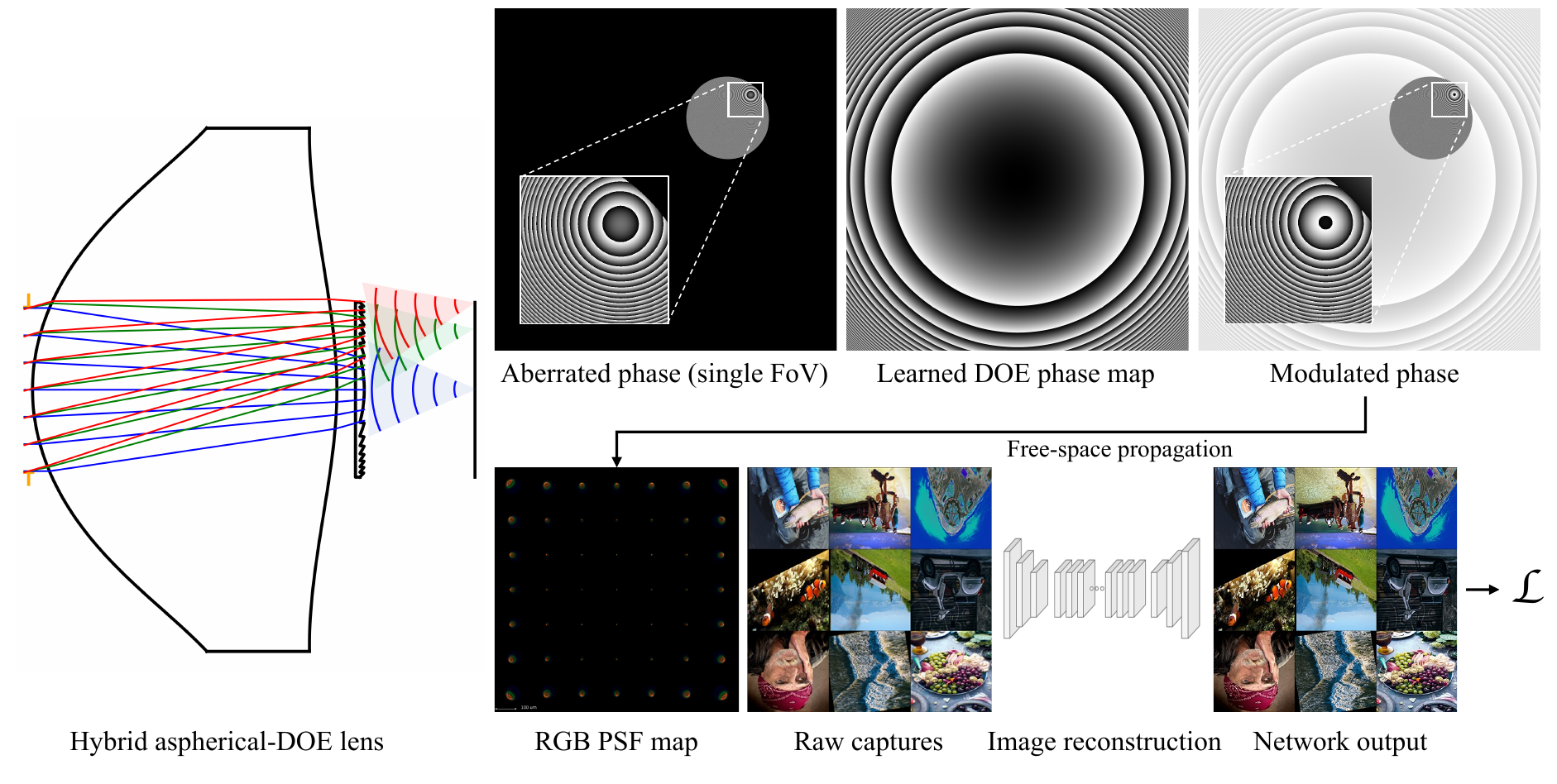}
    \caption{A differentiable ray-wave model is proposed, enabling accurate simulation of both refractive aberrations and diffractive phase modulations. The ray-wave model initially calculates the complex wave field at the DOE plane using coherent ray-tracing. The aberrated wave field encodes the amplitude and phase aberrations introduced by the refractive lens. Subsequently, the wave field is modulated by the DOE phase profile and propagated to the sensor image plane for PSF calculation. The sensor-captured images are simulated with full FoV RGB PSFs and then fed into the downstream image reconstruction network. The pipeline is fully differentiable, enabling gradient backpropagation for end-to-end design of both the optics and the neural network.}
    \label{fig:pipeline}
\end{figure*}

\section{Methods}

\subsection{Differentiable Ray-Wave Imaging Model}

Considering a hybrid refractive-diffractive lens with the DOE at the last surface, where diffraction mainly occurs on the DOE surface (see e.g., Fig.~\ref{fig:pipeline}). We propose a differentiable ray-wave method to accurately model the optical aberrations and diffractive phase modulation. The ray-wave model consists of two parts: coherent ray tracing to calculate the aberrated complex wave field at the DOE plane, followed by DOE phase modulation and free-space wave propagation to the sensor plane, where the PSF is calculated.

\subsubsection{Coherent Ray Tracing}

Monte Carlo ray tracing~\cite{lafortune1996mathematical,li2018differentiable,wang2022differentiable} is a widely used method to simulate light propagation by modeling optical light as a group of rays. In the coherent ray tracing stage, we sequentially calculate the ray intersection $\mathcal{I}$ and refraction $\mathcal{R}$ at each lens surface $\mathcal{S}$ for each ray, recording the position $\mathbf{o}$, direction $\mathbf{d}$, and phase $\phi$. The ray tracing process through a refractive lens can be represented as:
\begin{equation}
    \left\{
    \begin{array}{l}
        \mathcal{I}_{n}(\mathcal{S}_{n}): (\mathbf{o}^{n-1}, \mathbf{d}^{n-1}, \phi^{n-1}, \lambda) \mapsto (\mathbf{o}^{n}, \mathbf{d}^{n-1}, \phi^{n}, \lambda),\\[1.5ex] 
        \mathcal{R}_{n}(\mathcal{S}_{n}): (\mathbf{o}^{n}, \mathbf{d}^{n-1}, \phi^{n}, \lambda) \mapsto (\mathbf{o}^{n}, \mathbf{d}^{n}, \phi^{n}, \lambda),\\[1.5ex] 
        (\mathbf{o}, \mathbf{d}, \phi, \lambda) = (\mathcal{R}_{N}\mathcal{I}_{N}\ldots\mathcal{R}_{1}\mathcal{I}_{1})(\mathbf{o}^{0}, \mathbf{d}^{0}, \phi^{0}, \lambda),
    \end{array}
    \right.
\end{equation}
where $\lambda$ is the wavelength of the light, and $N$ is the number of lens surfaces. The phase change $\phi$ is calculated by the optical path difference. We assume the point light source is coherent, and all rays have the same initial phase.
The intersection $\mathcal{I}$ can be solved by Newton's method and the refraction $\mathcal{R}$ can be solved by Snell's law, which have been described in detail in existing works~\cite{wang2022differentiable,yang2023curriculum,chen2021optical}. Double-precision arithmetic is used to address the precision problem during the phase calculation, as the wavelength of optical rays is much smaller than the physical size of the optical lens.

After tracing rays through the refractive lens and reaching the DOE surface, we can calculate the complex wave field by coherent superposition. The wave field before the DOE surface can be represented as
\begin{equation}
    \mathbf{U}_{\text{DOE}^{-}} = \sum_{i=1}^{spp} u_{i} \exp{\left(j\phi_{i}\right)} \cos\left<\mathbf{d}_i, \mathbf{n}\right>,
    \label{eq:wave_field}
\end{equation}
where $u_{i}$ is the amplitude and $\phi_{i}$ is the phase of the $i$-th optical ray. $\mathbf{n}$ is the normal vector of the DOE surface, and $\cos\left<\mathbf{d}_i, \mathbf{n}\right>$ represents the obliquity factor in the Rayleigh-Sommerfeld theorem~\cite{goodman2005introduction}. $j$ is the imaginary unit, and $spp$ is the number of optical rays sampled from each point source, which is set to $10^{6}$ in our experiments.

Energy decay due to Fresnel transmission~\cite{born2013principles} is often ignored at the geometric lens design stage, and dealt with in a separate design step for optical coatings~\cite{baumeister2004optical}. In order to allow direct comparisons with existing systems we also ignore the energy loss along a ray; therefore, the amplitude $u_{i}$ of each ray is identical and set to 1. The complex amplitude of each optical ray is assigned to its surrounding 4 pixels with weights determined by the distance to each pixel point. This inverse bilinear interpolation enables gradient calculation when converting a group of rays to a wave field and also reduces the aliasing problem caused by sub-pixel phase shifts~\cite{mout2018ray}. The wave field $\mathbf{U}_{\text{DOE}^{-}}$ records both the amplitude and phase aberrations introduced by the imperfections of the refractive lens. An example of the phase of a single-FoV wave field before the DOE surface is shown in Fig.~\ref{fig:pipeline}, the majority of the phase is undefined (black) because light is converged to a small region by the refractive lens.

\subsubsection{DOE Phase Modulation}

The DOE surface introduces a phase modulation to the wave field passing through it, as illustrated in Fig.~\ref{fig:pipeline}. Based on scalar diffraction theory~\cite{goodman2005introduction} and using Kirchhoff boundary conditions~\cite{born2013principles}, the phase change introduced by the DOE can be expressed as
\begin{equation}
    \mathbf{U}_{\text{DOE}^{+}} = \mathbf{U}_{\text{DOE}^{-}} \exp{\left(j \frac{2\pi}{\lambda} \left(n_{\lambda} - 1 \right)h (x, y) \right)},
    \label{eq:phase_profile}
\end{equation}
where $h(x,y)$ is the 2D height map of the DOE surface, and $n_{\lambda}$ is the refractive index of the DOE substrate material at the wavelength $\lambda$. For compatibility with Zemax, the DOE design is internally represented as the phase $\phi_0$ at the nominal design wavelength $\lambda_{0}$, and then remapped to the corresponding phases of other wavelengths during simulation. Given that 
\begin{equation}
    \phi_{0} = \frac{2\pi}{\lambda_{0}} (n_0-1) h(x, y),
    \label{eq:phi0}
\end{equation}
by substituting Eq.~\eqref{eq:phi0} into Eq.~\eqref{eq:phase_profile}, the wave field after DOE modulation can be expressed as
\begin{equation}
    \mathbf{U}_{\text{DOE}^{+}} = \mathbf{U}_{\text{DOE}^{-}} \exp{\left(j\frac{n_{\lambda} - 1}{n_0 - 1}\frac{\lambda_{0}}{\lambda}\phi_0\right)}.
    \label{eq:phase_modulation2}
\end{equation}
In our experiments, $\lambda_{0}$ is set to 0.55~$\mu$m and $\phi_0$ is the optimizable parameter. In Eq.~\eqref{eq:phase_modulation2}, the DOE phase map $\phi_{0}$ can be a discontinuous function to model the multi-level DOE features, eliminating the continuous condition in existing methods~\cite{fischer2000optical,zhu2023metalens}.

\subsubsection{PSF Calculation}

The modulated wave field then propagates to the sensor plane. This free-space propagation can be calculated using the angular spectrum method~\cite{goodman2005introduction}, i.e.,
\begin{equation}
    \mathbf{U}_{\text{Sensor}} = \mathcal{F}^{-1}(\mathcal{F}(\mathbf{U}_{\text{DOE}^{+}}) \mathbf{H}),
\end{equation}
where $\mathbf{H}$ is the transfer function, while $\mathcal{F}$ and $\mathcal{F}^{-1}$ respectively represent the Fourier transform and its inverse. The Nyquist sampling criterion~\cite{mehrabkhani2017rayleigh,goodman2005introduction} requires a small sampling step of the wave field $\mathbf{U}_{\text{DOE}^{+}}$ for accurate calculation. Therefore, in Eq.~\eqref{eq:wave_field}, we choose a sampling resolution of an integer multiple (typically twice) of the DOE feature size and upsample the DOE phase map to the same resolution as the wave field. Specifically, for the lens example in Fig.~\ref{fig:pipeline}, the DOE is defined over a 3 mm $\times$ 3 mm area with a 1 $\mu$m sampling resolution. Hence, we sample a 6,000~$\times$~6,000 wave field in Eq.~\eqref{eq:wave_field} and upsample the DOE phase map to the same resolution. During the free-space wave propagation, we also pad the wave field with zeros to twice the physical size and resolution to cover off-axis waves~\cite{matsushima2010shifted}, corresponding to a 12,000~$\times$~12,000 resolution.

The PSF of the entire system is calculated by squaring the amplitude of the wave field at the sensor image plane, which can be represented as
\begin{equation}
    \mathbf{PSF} = |\mathbf{U}_{\text{Sensor}}|^2.
    \label{eq:psf}
\end{equation}
On the sensor image plane, the sampling resolution is determined by the camera sensor used, which is typically smaller than the wave field $\mathbf{U}_{\text{Sensor}}$. Therefore, an undersampling of the intensity distribution is performed. Subsequently, we crop the valid region, determined by the perspective relation of the refractive lens, to obtain a smaller PSF of the hybrid lens. 

We encourage the readers to refer to existing works~\cite{goodman2005introduction,born2013principles,mout2018ray,wang2022differentiable,yang2023curriculum,chen2021optical} for more details about coherent ray tracing and wave propagation methods. Our experimental code will also be released in the future to assist in understanding the proposed method.

\subsection{Mixed Precision End-to-End Optical Design}

A mixed-precision training method is proposed to bridge the double-precision optical simulation with single-precision network training. Specifically, we observe that the precision challenge primarily arises from phase calculations during coherent ray tracing and wave propagation, while the intensity PSF is less sensitive to precision. To address this, we introduce a differentiable PyTorch autograd function that converts the double-precision PSF to single-precision after Eq.~\eqref{eq:psf} for image simulation and network training. In the backpropagation phase, the single-precision PSF gradients from the network are converted back to double-precision and continue backpropagating through the optics.

We build the differentiable ray-wave model on top of an open-source ray tracer, DeepLens~\cite{yang2023curriculum,wang2022differentiable}, using the PyTorch framework. The memory consumption of the proposed differentiable ray-wave model is high due to million-scale ray tracing, high-resolution wave propagation, and downstream network processing. Specifically, for the hybrid aspherical-DOE lens (shown in Fig.~\ref{fig:pipeline}) with given experimental settings (10$^6$ rays, 6,000 $\times$ 6,000 field), the single FoV RGB PSF calculation and backpropagation require approximately 35~GB of GPU memory, which gives a theoretical lower bound of GPU memory requirement. The memory consumption for multi-FoV PSF calculation and end-to-end lens design can be reduced by several strategies, such as multi-GPU parallelization, gradient checkpointing, patch backpropagation, and adjoint rendering~\cite{yang2023curriculum,vicini2021path,teh2022adjoint,nimier2020radiative}. In our end-to-end hybrid lens design experiments, we calculate 10$\times$10 RGB PSFs for accurate simulation of the hybrid lens system. We choose three wavelengths to represent the broadband for each color channel and randomly select one in each iteration to calculate the PSF during the training. The PSFs are then convolved with the input image batch for output image simulation and downstream network training, which has been described in detail in existing end-to-end lens design works~\cite{sitzmann2018end,tseng2021neural,cote2023differentiable}. See more experimental details in the following sections and Supplementary Materials.

%% file: 4_verification.tex
\section{Model Accuracy}

\begin{figure}[!htp]
    \centering
    \includegraphics[width=0.7\textwidth]{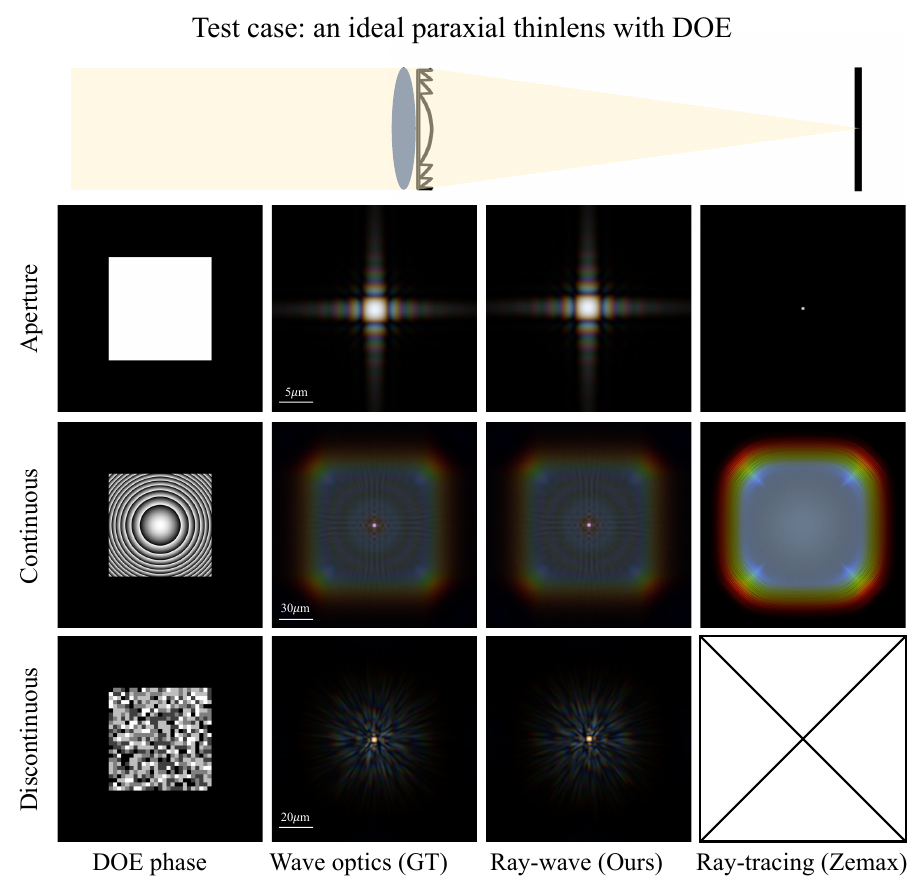} 
    \caption{PSF simulation of different optical models. An ideal paraxial thin lens with a DOE is used for testing. For this specific case, wave optics gives accurate simulation results~\cite{goodman2005introduction}, therefore serving as the ground truth. Different DOE phase maps are studied, our ray-wave model gives identical results with the ground truth, while the ray tracing model fails to address the real diffractive phenomena and can not function for discontinuous phase maps.}
    \label{fig:validation}
\end{figure}

The ray-tracing and wave propagation described above are individually well understood and validated, however, their combination to model hybrid systems is new and must be evaluated. Unfortunately, there is no ``ground truth'' reference for complex hybrid systems that could be compared against. We therefore turn to a simplified optical system that can be simulated with our hybrid model, but also entirely with scalar diffraction theory as a ground truth solution. The optical system (Fig.~\ref{fig:validation}) consists of an ideal (i.e., un-aberrated) thin lens and a DOE. The thin lens model can be easily simulated with both ray-tracing and scalar diffraction theory~\cite{goodman2005introduction}, allowing us to compare our hybrid model to the scalar diffraction solution as well as a hybrid simulation provided by the commercial Zemax lens design package. Zemax uses a ray-tracing model with an additional ray-bending term based on the grating equation to approximate diffraction~\cite{ZemaxManual,yu2011light}.

The specifications of the test are as follows. The lens consists of a thin lens ($f$ = 100~mm) and a DOE attached to it. Three different DOE phase maps are studied: (1) a constant phase with a square aperture to simulate regular aperture diffraction; (2) a continuous phase map to simulate an ideal Kinoform diffractive surface~\cite{jordan1970kinoform}; and (3) a discontinuous phase map to simulate both the multi-level fabrication process and discontinuous DOEs in existing works~\cite{sun2020learning,shi2022seeing}. The simulation results indicate that our ray-wave model yields virtually identical results to the ground truth, while the ray tracing model fails. Specifically, the edge diffraction effect of the aperture mask is entirely ignored by the ray tracing model, and the central bright spot of the continuous phase map is caused by the interference of the diffracted light, which is also not captured by the ray tracing model. Moreover, for multi-level discontinuous phase maps, the ray tracing model cannot function because it requires a continuous condition to calculate the gradient of the local phase~\cite{fischer2000optical}. More PSF simulation results are presented in the Supplementary Material.

A comprehensive comparison with other existing simulation methods for refraction and diffraction simulation is presented in Table~\ref{tab:model_compare}. Specifically, we primarily consider imaging models that are or can be easily designed to be differentiable. Chen et al.~\cite{chen2021optical} proposed a ray tracing model to simulate the aperture diffraction in a refractive lens; however, this exit-pupil diffraction method cannot function for diffractive optical elements. Zhu et al.~\cite{zhu2023metalens} introduced a wave-ray model, which converts the wavefront after phase modulation into a group of optical rays for ray tracing. Similar to Zemax~\cite{ZemaxManual}, this model also relies on the gradient calculation of the phase map~\cite{yu2011light}, making it inaccurate and unable to function for binary and discontinuous phase maps. A detailed comparison is provided in the Supplementary Material.

\begin{table}[tp]
\centering
\small
\caption{Comparison of different hybrid refractive-diffractive lens simulation models. A detailed explanation is provided in the supplementary material.}
\label{tab:model_compare}
\renewcommand{\arraystretch}{2.0}
\resizebox{0.7\textwidth}{!}{
\begin{tabular}{l|c|c|c|c|c}
\hline
\multirow{1}{*}{} & Paraxial optics & Zemax~[\cite{ZemaxManual}] & Chen~\etal~[\cite{chen2021optical}] & Zhu~\etal~[\cite{zhu2023metalens}] & Ours \\ \hline
Optical model & Wave & Ray & Ray & Wave-ray & Ray-wave \\ \hline
Accuracy & \hfil\red    & \hfil\orange & \hfil\green   & \hfil\orange    & \hfil\green   \\ 
Optical aberration        & \hfil\red    & \hfil\green  & \hfil\green   & \hfil\green     & \hfil\green   \\
Edge diffraction          & \hfil\green  & \hfil\red    & \hfil\green   & \hfil\green     & \hfil\green   \\
\hline
Phase modulation          & \hfil\green  & \hfil\green  & \hfil\red     & \hfil\green     & \hfil\green   \\
Discontinuous phase       & \hfil\green  & \hfil\red    & \hfil\red     & \hfil\red       & \hfil\green   \\
\hline
Differentiable            & \hfil\green  & \hfil\green  & \hfil\orange  & \hfil\green     & \hfil\green   \\
End-to-end design         & \hfil\green  & \hfil\orange & \hfil\red     & \hfil\green     & \hfil\green   \\
\hline
\end{tabular}
}
\end{table}

%% file: 5_simu_results.tex
\begin{figure*}[ht]
    \centering
    \includegraphics[width=\textwidth]{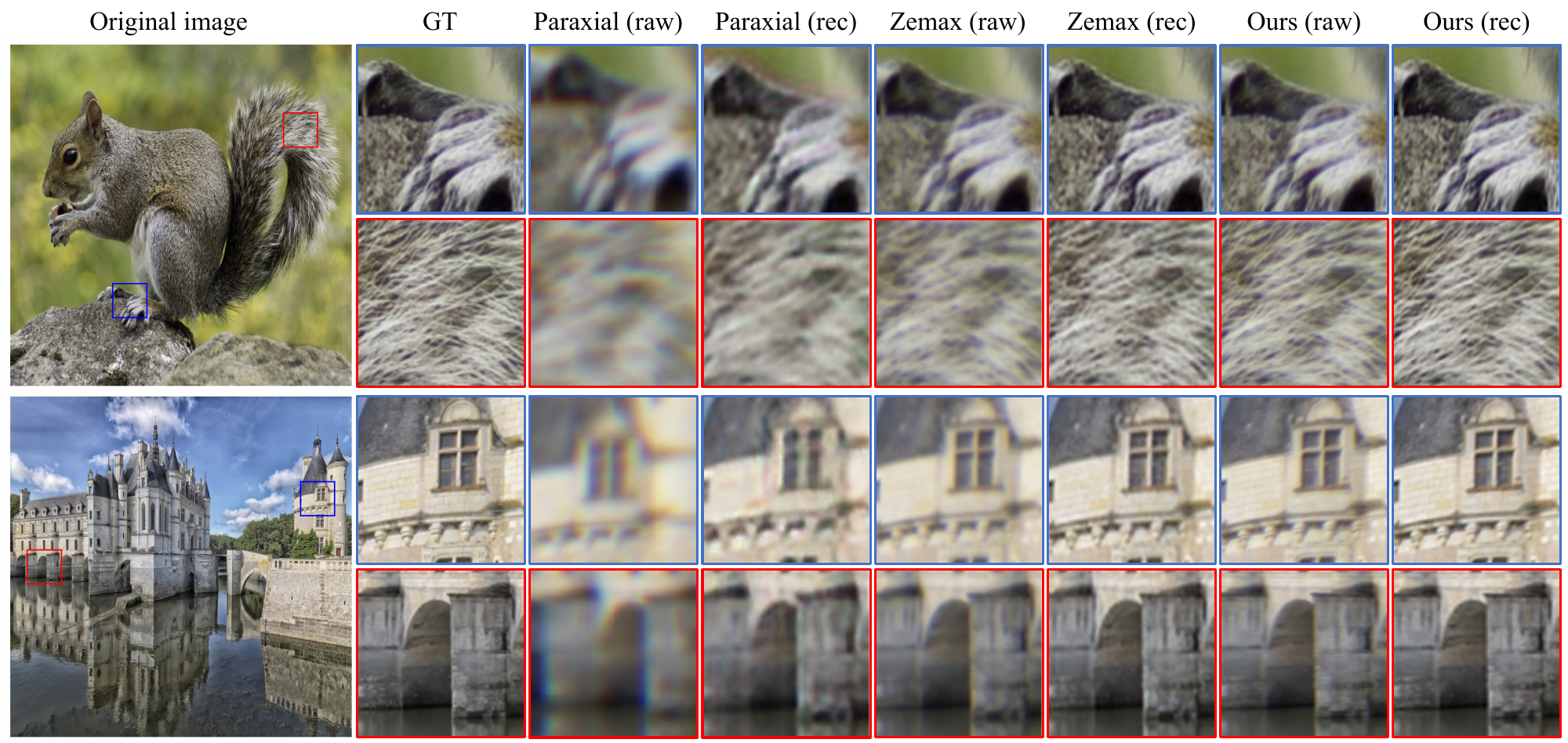}
    \caption{Comparison of image quality between different hybrid lens design approaches. The paraxial wave optics model fails to correct aberrations across the full FoV, resulting in degraded image quality in both raw simulations and network reconstructions. The ray tracing method in Zemax corrects aberrations but uses the RMS spot size as the objective, precluding end-to-end lens design. Our ray-wave model accurately simulates full-FoV PSFs and is fully differentiable, enabling end-to-end co-design of the hybrid lens and the network, thus achieving the best reconstruction quality. Raw simulated images and corresponding reconstructed images are shown for each design approach, with zoomed-in patches highlighting the differences in image quality across the FoV.}
    \label{fig:cooke_img}
\end{figure*}

\section{End-to-end hybrid lens design}

\begin{table*}[t]
\small
\centering
\caption{Performance comparison of different hybrid lens designs by simulation. The PSNR, SSIM, and 1-LPIPS matrix are calculated for both simulated (``raw'') and reconstructed (``rec'') images.}
\label{tab:cooke_score}
\renewcommand{\arraystretch}{1.2} 
\resizebox{\textwidth}{!}{
\begin{tabular}{l|c|c|c}
\hline
Lens design & RMS radius (on-axis/off-axis/avg) $\downarrow$ & PSNR/SSIM/1-LPIPS (raw) $\uparrow$ & PSNR/SSIM/1-LPIPS (rec) $\uparrow$   \\
\hline
Paraxial wave optics            & 16.1/55.0/32.0         & 17.7/0.520/0.443          & 25.041/0.688/0.569             \\
Ray tracing (Zemax)             & \textbf{7.6/19.6/10.1} & 24.8/0.728/0.706          & 36.462/0.963/0.924             \\
Ray-wave model (ours)           & 10.7/28.4/15.4         & \textbf{27.1/0.795/0.713} & \textbf{39.9/0.982/0.963}    \\
\hline
\end{tabular}
}
\end{table*}

\begin{figure}[!htp]
    \centering
    \includegraphics[width=0.6\textwidth]{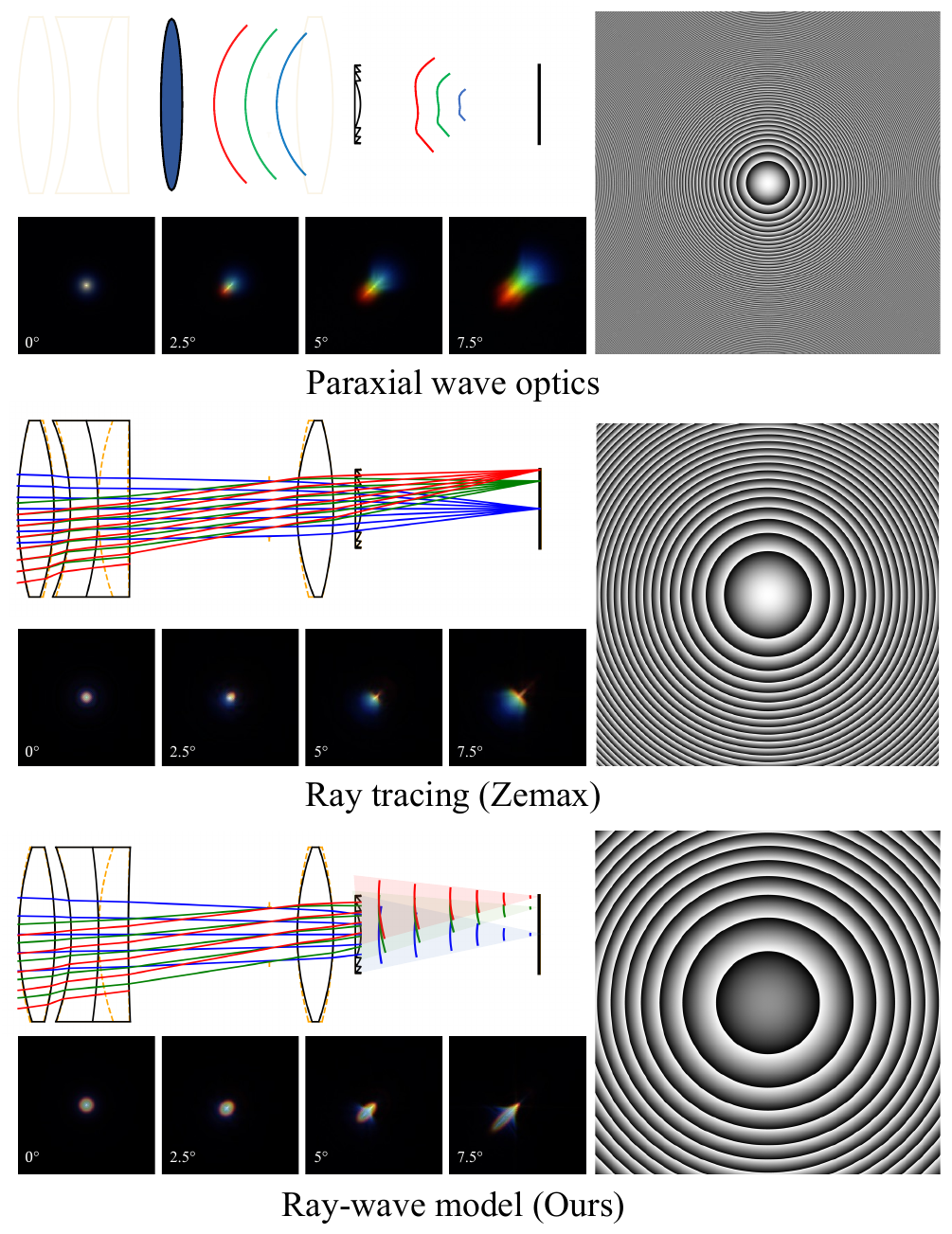}
    \caption{Hybrid lens designs using different optical models and optimization methods. Top: achromatic DOE design using the paraxial optical model, minimizing paraxial chromatic aberration. Middle: hybrid lens design using a ray-tracing model, optimizing the RMS spot size in Zemax. Bottom: hybrid lens design using the proposed ray-wave model, optimizing the final image quality. PSFs at different FoVs are calculated with the proposed ray-wave model, shown in the log scale.
    }
    \label{fig:cooke_psf}
\end{figure}

We first conduct an end-to-end hybrid lens design by simulation. A compound refractive lens, a DOE, and an image reconstruction network are jointly optimized. The refractive lens has a total length of 75~mm, a focal length of 47~mm, a sensor size of 8 mm $\times$ 8 mm, a diagonal FoV of 14~$^\circ$ at F/4. The refractive lens exhibits strong chromatic aberrations, being a good candidate for DOE aberration correction. The DOE is placed between the refractive lens and the image sensor, with a physical size of 8 mm $\times$ 8 mm. The designed DOE phase map $\phi_0$ is parameterized as $\phi_0 = \sum_{i=1}^{k} \alpha_{i} \rho^{2i}$, where $\rho$ is the normalized radial distance from the center of the DOE, and $\alpha_{i}$ is the $i$-th even-polynomial coefficient. In our experiments, we set $k=4$. Both the aberrated wave field $\mathbf{U}_{\text{DOE}^{-}}$ and the DOE phase map have a resolution of 5,000 $\times$ 5,000. During the end-to-end lens design process, the refractive lens, DOE, and an imaging reconstruction network are simultaneously optimized to achieve the best image output. The sensor has a resolution of 2,000 $\times$ 2,000, corresponding to a pixel pitch of 4 $\mu$m. We sample 10 $\times$ 10 RGB PSFs for full-resolution image simulation. For image reconstruction, we adopt NAFNet~\cite{chen2022simple}, which demonstrates outstanding image deblurring performance and fast convergence. The image reconstruction loss is used for end-to-end lens design to learn the optimal optics-network joint system, which can be represented as:
\begin{equation}
    \mathcal{L} = \mathcal{L}_{\mathrm{pixel}} \left(\mathcal{N}\left(\mathbf{PSF} \ast \mathbf{I}\right), \mathbf{I}\right) + \alpha\mathcal{L}_{\mathrm{percep}}\left(\mathcal{N}\left(\mathbf{PSF} \ast \mathbf{I}\right), \mathbf{I} \right),
\end{equation}
where $\mathbf{I}$ is the input object image, which is also used as the ground truth since we aim to optimize for the best imaging performance. $\mathcal{N}$ represents the reconstruction network, $\mathcal{L}_{\mathrm{pixel}}$ is the pixel-wise loss, $\mathcal{L}_{\mathrm{percep}}$ is the perceptual loss, and $\alpha$ is the weight of the perceptual loss. In our experiments, we use mean squared error loss as the pixel-wise loss and VGG loss~\cite{johnson2016perceptual} as the perceptual loss with $\alpha=0.1$. DIV2K dataset~\cite{agustsson2017ntire} is used for training and testing. The end-to-end training runs for 50~epochs, then we fix the hybrid lens and fine-tune the image reconstruction network with sensor noise for 50 epochs. 

Paraxial wave optics and ray tracing models are used for comparison: (1) The paraxial wave model (Fig.~\ref{fig:cooke_psf}) represents the compound refractive lens as a phase plate. We pre-calculate the on-axis chromatic aberrations of the refractive lens and design the DOE for achromatic purposes~\cite{wang2016chromatic,chen2018broadband}. Specifically, we use the Fresnel phase DOE and analytically solve it to reduce the chromatic aberration at different wavelengths. (2) For the ray tracing model (Fig.~\ref{fig:cooke_psf}), we load the hybrid lens into Zemax and jointly optimize the refractive lens with the DOE, using the \textit{``Binary2''} surface with the same polynomial order as ours, operated by an experienced optical designer. We then load the optimized hybrid lens into our code and continue optimizing the optics to close the gap between our simulation and Zemax, with RMS spot size used as the optimization objective.

Three final designs are presented in Fig.~\ref{fig:cooke_psf}. The orange dashed lines indicate the original refractive lens, and the DOE phase map is shown on the right. The log-scale PSFs at 4 FoVs from on-axis to full-FoV are calculated using our proposed model, which has been proven to yield accurate simulations in the previous section. The paraxial wave optics model fails to consider the off-axis aberrations, thus leaving significant chromatic aberrations at large FoVs. For both ray optics and our ray-wave model, the PSFs at all FoVs are optimized to be small. The RMS spot size is calculated with the ray tracing model and listed in Table~\ref{tab:cooke_score}. The Zemax design performs the best in terms of RMS spot size, which is the loss it was designed for.

We further evaluate the three lenses based on high-level imaging quality. For each lens, we simulate sensor-captured images using our proposed ray-wave model and train an image reconstruction network for each different lens to achieve the best output quality. In this network fine-tuning stage, 40$\times$40 PSFs are used for accurate image simulation, and Gaussian noise is added to simulate sensor noise. As presented in Table~\ref{tab:cooke_score}, PSNR, SSIM, and 1-LPIPS metrics are calculated for the evaluation of both simulated images (``raw'') and reconstructed images (``rec''). Our end-to-end designed hybrid lens using the ray-wave model successfully outperforms the other two designs in terms of image quality, since we directly optimize the final image, whereas the other two models cannot perform end-to-end design. Example ``raw'' and ``rec'' images are provided in Fig.~\ref{fig:cooke_img}, with zoomed image patches. More results and evaluations are provided in the Supplementary Material.

%% file: 6_real_results.tex
\begin{figure*}[tp]
    \centering
    \includegraphics[width=\textwidth]{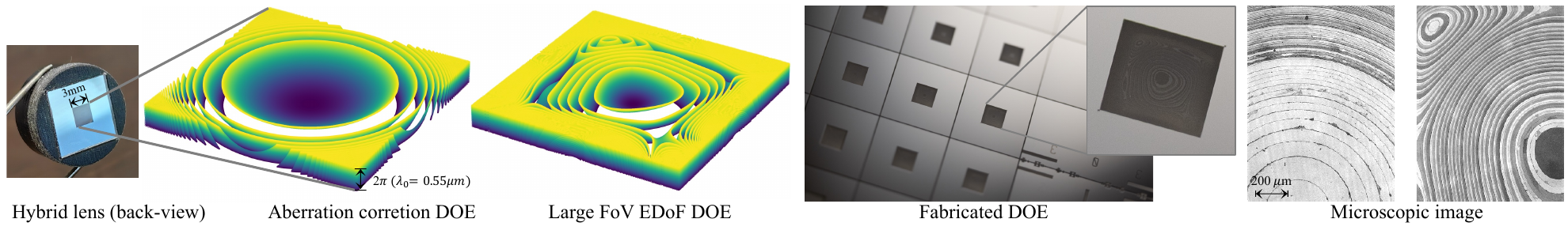}
    \caption{Designed DOE phase map (left) and fabricated samples (right) for computational aberration correction and large FoV EDoF imaging. Microscopic images are taken by a Nikon objective, 5X/0.13 TI WD 9.3 427041.}
    \label{fig:doe_zoom}
\end{figure*}

\section{Hybrid aspheric-DOE lens prototype}

To evaluate the proposed method with real-world experiments, we designed and built a large FoV compact aspherical-DOE hybrid lens prototype. Since we lack the fabrication capabilities for refractive lenses, we utilize an off-the-shelf aspherical lens and use our hybrid framework to only optimize the DOE, taking into account the real aberrations of the refractive lens. An aspherical lens with known surface data (Optolife Optics, China.) with a focal length of 7.5~mm is chosen for evaluation, as shown in Fig.~\ref{fig:teaser}. The image sensor (OmniVision OV2710) has a pixel pitch of 3~$\mu$m~$\times$~3~$\mu$m, and the center 1,000 $\times$ 1,000 imaging region is used for the experiments. The diagonal FoV of the refractive lens is 35~$^\circ$ at an effective F/2.2, leading to significant off-axis optical aberrations. Various fabrication techniques can be employed to fabricate DOEs, including lithography+etching~\cite{shi2022seeing,fu2022diffractive,zheng2023hexagonal}, lithography+deposition~\cite{amata2023additive}, and lithography+nanoimprint~\cite{fu2021etch}. The DOE has a feature size of 1~$\mu$m~$\times$~1~$\mu$m and a physical size of 3~mm~$\times$~3~mm. To best preserve the edge features and micro-profiles, we choose to fabricate the DOE with high spatial resolution in a 16-level structure. The lens and DOE are then assembled with a 3D-printed mount for the final prototype. 

Two applications are demonstrated to evaluate the performance of our hybrid aspherical-DOE lens: (1) computational aberration correction and (2) large FoV and EDoF imaging. Specifically, existing diffractive EDoF works~\cite{li2023extended,pinilla2022hybrid,seong2023e2e} idealize the refractive lens as a paraxial thin lens, ignoring aberrations and functioning only for a small FoV. Our proposed model accurately simulates optical aberrations, enabling large FoV EDoF imaging. Moreover, since we do not optimize the refractive lens part, the aberrated wavefield at the DOE surface can be pre-calculated and stored, making the memory and time consumption for the single FoV PSF simulation the same as the paraxial optical model. All experimental settings are kept consistent (see Supplementary Material for more details).

\subsection{Computational Aberration Correction}

\begin{figure*}[ht]
    \centering
    \includegraphics[width=\textwidth]{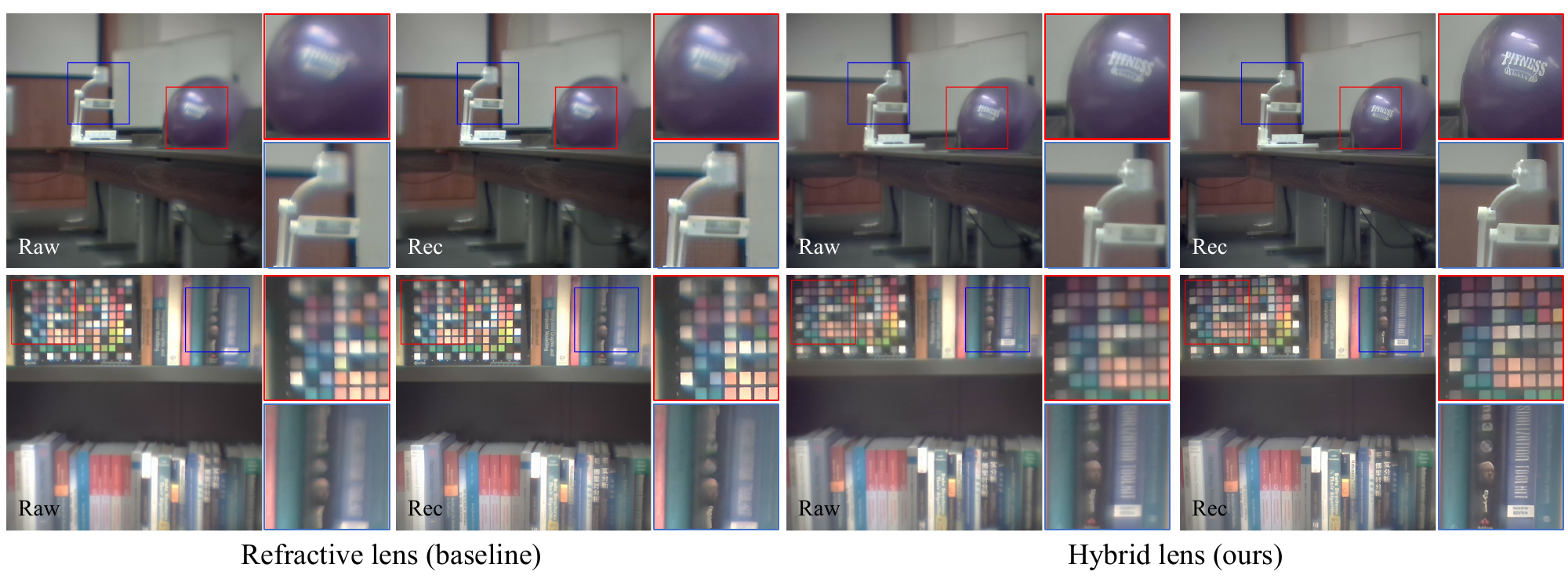}
    \caption{Real-world image quality comparison between the refractive lens (baseline) and the hybrid lens (ours). Our designed and fabricated DOE corrects the aberrations of the refractive lens and achieves better image quality, especially for the off-axis image regions where aberrations are more significant.}
    \label{fig:a951_img}
\end{figure*}

An aberration correction DOE is end-to-end designed with the reconstruction network. The same DOE parameterization is adopted for computational aberration correction. The optimized DOE phase map and a microscope image of the fabricated DOE are presented in Fig.~\ref{fig:doe_zoom}. For comparison, the original aspherical lens with a blank DOE (to maintain the same aperture) is selected as the baseline experiment. Quantitative evaluation scores are measured on the simulated dataset, as presented in Table~\ref{tab:asphe_score}. In Fig.~\ref{fig:a951_img}, both the raw captured (``raw'') and reconstructed (``rec'') images in the real world are shown for the hybrid lens and the refractive lens. The results demonstrate that the hybrid lens can successfully correct the refractive aberrations and achieve better image quality, especially for the off-axis image regions where aberrations are more significant. More results and experimental details are provided in the Supplementary Material. Moreover, inserting the DOE into an existing refractive lens at the back does not increase the form factor, showing great potential for cameras with highly constrained physical size, such as cellphone cameras.

\begin{table}[t]
\small
\centering
\caption{Performance comparison between refractive and hybrid lens designs on the simulated dataset.}
\label{tab:asphe_score}
\renewcommand{\arraystretch}{1.5} 
\resizebox{0.7\columnwidth}{!}{
\begin{tabular}{l|c|c}
\hline
Lens & PSNR/SSIM/1-LPIPS (raw) & PSNR/SSIM/1-LPIPS (rec)\\
\hline
Refractive lens & 24.4/0.702/0.677 & 28.8/0.835/0.746 \\
Hybrid lens (ours) & 24.7/0.733/0.707 & \textbf{32.5/0.931/0.880} \\
\hline
\end{tabular}
}
\end{table}

\subsection{Aberration-Aware Large Field-of-View Extended-Depth-of-Field Imaging}

\begin{figure*}[ht]
    \centering
    \includegraphics[width=\textwidth]{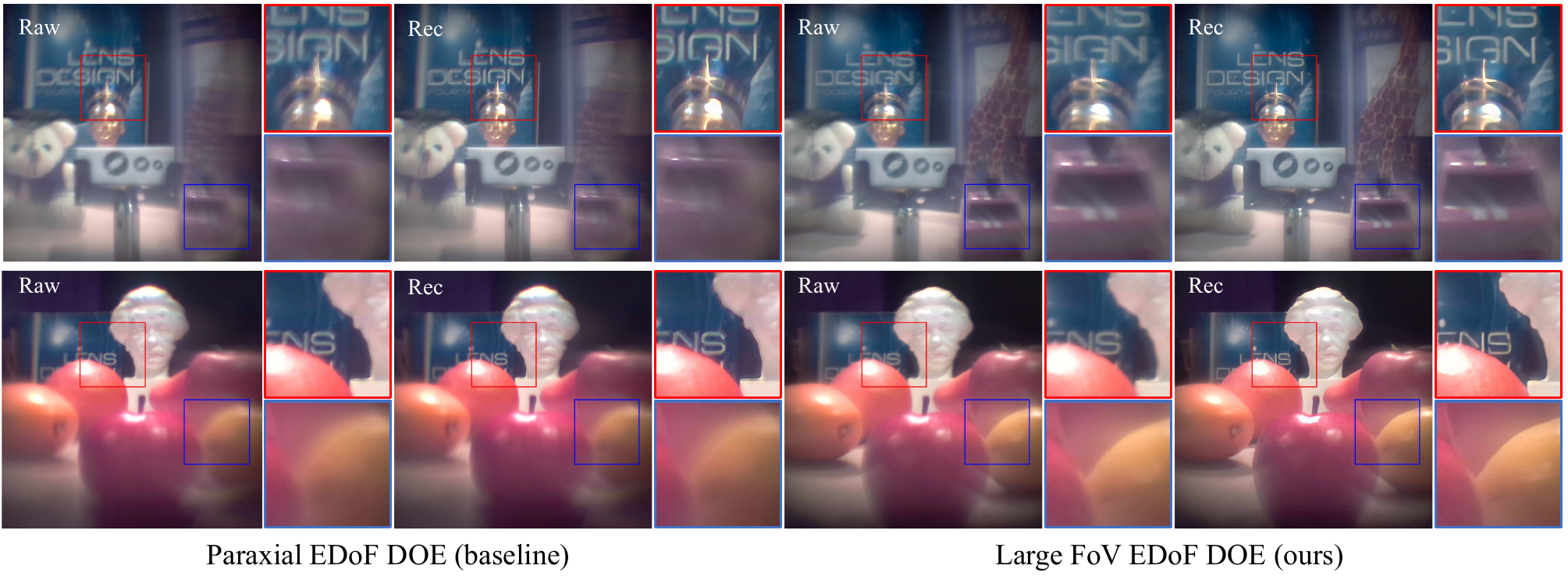}
    \caption{Real-world image quality comparison of extended depth-of-field imaging between the DOEs design with the paraxial optical model (baseline) and the proposed ray-wave model (ours). Hybrid lens with our designed and fabricated DOE achieves better image quality over a large FoV and a large depth of field, particularly in the off-axis regions which are neglected during the design using the paraxial optical model.}
    \label{fig:edof_img}
\end{figure*}

An EDoF DOE is end-to-end designed with the proposed ray-wave model to image clearly from 20~cm to 10~m with a diagonal FOV of 35$^{\circ}$, named ``large FoV DOE''. The DOE is parameterized as $\phi_0 = \sum_{i=2}^{k} \beta_{i} \rho^{i}$. We set $k=7$ in our experiments. The odd polynomials provide the EDoF capability, while the even polynomials correct optical aberrations. With accurate simulation of optical aberrations, the designed DOE can find the best phase map in the presence of optical aberrations, thus improving image quality and practical performance in the real world. The optimized DOE phase map and a microscope image of the fabricated DOE are in Fig.~\ref{fig:doe_zoom}.

For comparison, a DOE is end-to-end designed using the paraxial optical model, referred to as the ``paraxial DOE,'' as presented in existing works~\cite{pinilla2022hybrid,sitzmann2018end}. Subsequently, we evaluate the simulated images of the two DOEs using the proposed ray-wave model and reconstruct the simulated images with their respective reconstruction networks. Quantitative results measured on the simulated dataset are presented in Table~\ref{tab:edof_score}, with qualitative results on the simulated dataset shown in Fig.~\ref{fig:edof_img}. We then fabricated both DOEs for real-world experiments, and the results are depicted in Fig.~\ref{fig:edof_img}. The results demonstrate that the hybrid lens can image clearly over a large depth-of-field and wide field-of-view (FoV). Notably, for off-axis image regions, the DOE designed with our ray-wave model can maintain excellent image quality, while the DOE designed with the paraxial optical model fails to image clearly. Additional results and experimental details are provided in the supplementary material.

\begin{table}[t]
\centering
\small
\caption{Performance comparison of EDoF DOE designed with different imaging models.}
\label{tab:edof_score}
\renewcommand{\arraystretch}{1.5} 
\begin{tabular}{c|l|c|c}
\hline
DOE &depth & PSNR/SSIM/1-LPIPS (raw) & PSNR/SSIM/1-LPIPS (rec)\\ \hline
\multirow{3}{*}{\rotatebox{90}{Baseline}} & 20~cm & 18.7/0.497/0.531 & 16.1/0.341/0.509 \\
 & 30~cm & 19.0/0.539/0.473 & 18.7/0.520/0.437 \\
 & 10~m & 18.9/0.523/0.517 & 18.6/0.502/0.493 \\ \hline
\multirow{3}{*}{\rotatebox{90}{Ours}} & 20~cm & 21.5/0.575/0.587 & \textbf{27.5/0.821/0.782} \\
 & 30~cm & 22.4/0.659/0.635 & \textbf{28.9/0.869/0.842} \\
 & 10~m & 21.7/0.598/0.524 & \textbf{27.4/0.818/0.787} \\ \hline
\end{tabular}
\end{table}

%% file: 7_conclusion.tex
\section{Conclusion and Discussion}

In this paper, we propose a differentiable ray-wave model for hybrid refractive-diffractive optical systems. The proposed model can accurately simulate both optical aberrations and diffractive phase modulation while enabling gradient backpropagation for end-to-end co-design of hybrid lenses and neural networks. We validate the simulation accuracy and employ the ray-wave model to jointly design a hybrid lens consisting of a compound refractive lens, a DOE, and an image reconstruction network. We compare our results with existing methods and commercial software, such as Zemax, demonstrating the effectiveness of our approach. Furthermore, we demonstrate the practical applicability of our proposed model through an aspherical-DOE lens prototype for aberration correction and large field-of-view EDoF imaging. Both simulations and real experiments show that the proposed prototype achieves high-quality imaging performance, outperforming existing methods. 

The proposed model paves the way towards high-quality imaging systems with challenging optical requirements, such as compact cellphone lenses~\cite{reshef2021optic,williams2023three} with large aperture and wide field of view, clip-in diffractive filters for commercial interchangeable-lens cameras~\cite{JP2023128253A}, and augmented reality wave-guide~\cite{tseng2024neural,gopakumar2024full} for near-display applications, to name just a few. Placing the DOE away from the Fourier plane is beneficial to shrink its size while maintaining a large entrance pupil of the system. In addition, optimizing the DOE position to the sensor adds further flexibility to the optimization. Last but not least, one limitation of placing the DOE as the last optical component in the hybrid system is owing to the current difficulties in reverse conversion from wave to rays, requiring further improvements in the model.